\documentclass[aps,prd,nofootinbib,twocolumn,superscriptaddress,preprintnumbers]{revtex4}

\pdfoutput=1

\usepackage{hyperref}
\usepackage{epsfig}
\usepackage{setspace}
\usepackage{textcomp}
\usepackage{amsmath}
\usepackage{amssymb}
\usepackage{graphicx}
\usepackage{slashed}
\usepackage{multirow}

\usepackage{subfigure}

\usepackage{color}

\newcommand{\eg}{{\it e.g.}}

\newcommand{\be}{\begin{equation}}
\newcommand{\ee}{\end{equation}}
\newcommand{\br}{\begin{eqnarray}}
\newcommand{\bea}{\begin{eqnarray}}
\newcommand{\eea}{\end{eqnarray}}
\newcommand{\er}{\end{eqnarray}}
\newcommand{\ba}{\begin{array}}
\newcommand{\ea}{\end{array}}
\newcommand{\bi}{\begin{itemize}}
\newcommand{\ei}{\end{itemize}}
\newcommand{\bn}{\begin{enumerate}}
\newcommand{\en}{\end{enumerate}}
\newcommand{\bc}{\begin{center}}
\newcommand{\ec}{\end{center}}

\newcommand{\beq}{\begin{equation}}
\newcommand{\eeq}{\end{equation}}

\newcommand{\gsim}{\lower.7ex\hbox{$\;\stackrel{\textstyle>}{\sim}\;$}}
\newcommand{\lsim}{\lower.7ex\hbox{$\;\stackrel{\textstyle<}{\sim}\;$}}

\newcommand{\bs}{\begin{small}}
\newcommand{\es}{\end{small}}

\newcommand{\qui}{q_{{\scriptscriptstyle U}_{\!i}}}
\newcommand{\qdi}{q_{{\scriptscriptstyle D}_{\!i}}}

\newcommand{\eu}{e_{{\scriptscriptstyle U}}}
\newcommand{\ed}{e_{{\scriptscriptstyle D}}}

\def\mysection#1{{\bf #1.} }

\begin{document}

\title{Dark-Photon searches via Higgs-boson production at the LHC}

\author{Sanjoy Biswas}
\affiliation{KIAS, 85 Hoegi-ro, Dongdaemun-gu, 
Seoul 130-722, Republic of Korea}
\author{Emidio Gabrielli}
\affiliation{Dipart. di Fisica Teorica, Universit\`a di Trieste, Strada Costiera 11, I-34151 Trieste,  and INFN, Sezione di Trieste, Via Valerio 2, I-34127 Trieste, Italy}
\affiliation{NICPB, Ravala 10, 10143 Tallinn, Estonia}
\author{Matti Heikinheimo}
\affiliation{Helsinki Institute of Physics,
P.O.Box 64, FI-00014 University of Helsinki, Finland}
\author{Barbara Mele}
\affiliation{INFN, Sezione di Roma,  P. le A. Moro 2, I-00185 Rome, Italy}

\date{\today}
\begin{abstract}
Dark photons $\bar \gamma$ mediating long-range forces in a dark sector are predicted by various new physics scenarios, and are being intensively searched for 
in  experiments. We extend a previous study  of a new 
discovery process for dark photons proceedings via Higgs-boson production at the LHC. Thanks to the non-decoupling properties of the Higgs boson,   BR($H\to \gamma\bar \gamma$) values up to a few percent are possible for a massless dark photon, even for  heavy dark-sector scenarios. The corresponding 
 signature  consists (for a Higgs boson at rest) of a striking monochromatic photon 
 with 
energy $E_{\gamma}= m_H/2$, and similar amount of missing energy.
We perform a model independent analysis at the  LHC of both the gluon-fusion and VBF Higgs production mechanisms at 14~TeV, including parton-shower effects, and updating our previous parton-level analysis at 8 TeV in the gluon-fusion
channel by a more realistic background modeling.
We find that a $5\sigma$ sensitivity can be reached in the gluon-fusion channel for  
BR($H\to \gamma\bar \gamma)\simeq\,$0.1\% with an integrated luminosity of~$L\simeq 300\, {\rm fb}^{-1}$. The corresponding VBF reach is instead restricted  to~1\%.
 Such decay rates can be naturally obtained in 
dark-photon scenarios arising  from unbroken
$U(1)_F$ models explaining the origin and hierarchy of the  Yukawa couplings, strongly motivating the search for this exotic Higgs decay 
at the~LHC.
\end{abstract}
\preprint{HIP-2016-07/TH}
\maketitle

\section{Introduction}
Although long awaited, no conclusive signal of New Physics (NP) at the TeV scale showed up in  the first run of LHC at 7 and 8 TeV, and in the initial phase of  Run~2 at 13~TeV either. 
As a consequence, consents  are growing up around the idea that a new and unexplored {\it dark} (or {\it hidden}) sector, weakly coupled to the standard model (SM), is responsible for the observed dark matter (DM). The latter, which is five times more abundant in the Universe than ordinary baryonic matter~\cite{Ade:2013zuv}, remains still a mystery, with its constituents and detailed properties yet unknown. A dark sector could then have its internal structure and interactions, in complete agreement with present astroparticle and cosmological observations. 

It is also conceivable that a hidden sector could contain an extra long-range force 
mediator among the dark particles. The most simple 
example is provided by a new unbroken $U(1)$ gauge group, predicting a dark (or hidden) photon in its spectrum~\cite{Essig:2013lka}.
 Dark-photon scenarios have been extensively considered in 
the literature in the framework of NP extensions of the SM gauge
group \cite{Holdom:1985ag}-\cite{Falkowski:2014ffa}. 
  
In cosmology, dark photons may help to solve the small-scale structure formation problems. 
Massless dark photons interacting with dark matter \cite{Ackerman:mha}
 can lead to the formation of dark discs of galaxies \cite{Fan:2013tia}, analogously to the galaxy structure formation in the ordinary universe, or to the collisional behavior of dark matter in mergers of galaxies and galaxy clusters \cite{Heikinheimo:2015kra}.
In astroparticle physics, dark photons may induce the
Sommerfeld enhancement of DM annihilation cross section needed 
to explain the PAMELA-Fermi-AMS2 positron anomaly \cite{ArkaniHamed:2008qn}
, as well as assisting
light-DM annihilations to make asymmetric DM scenarios phenomenologically viable
\cite{Zurek:2013wia}. In some scenarios, 
massive dark photons  have also been considered as potential 
dark-matter candidates, with dedicated experiments looking for their direct detection in the mass range from a few eV up to 100 KeV~\cite{Essig:2013lka},\cite{An:2014twa}.

Most of present astrophysical and accelerator constraints
apply to {\it massive} dark photons, and can be evaded in the case of a {\it massless}
dark-photon scenario, allowing for potentially large dark-photon 
couplings in the dark sector. Indeed, in the massless case, {\it on-shell} dark photons can be fully decoupled from the SM quark and lepton sector \cite{Holdom:1985ag}, which is not true for the massive case due to the  potential tree-level mixing with ordinary photons
of massive dark photons. This property can lead to observable new signatures at colliders
for massless dark photons, provided there is a messenger sector letting the SM and dark sector communicate.

Recently, a massless dark photon scenario has been foreseen
in the framework of a theoretical proposal aimed to naturally solve the 
flavor hierarchy problem \cite{Gabrielli:2013jka}.
This model predicts a new Higgs-boson decay channel into a photon ($\gamma$) and a massless 
dark-photon ($\bar{\gamma}$) 
\bea
H\to \gamma \, \bar{\gamma}\, ,
\label{Hgdp}
\eea
which 
is induced at one-loop. The final $\bar{\gamma}$ gives rise to missing energy  and momentum  in the detector, leading to an exotic resonant mono-photon signature at the LHC. The latter features  
a distinctive photon transverse-momentum ($p_T^{\gamma}$) distribution 
peaked around $m_H/2$, same for the missing transverse-energy ($\slashed{E}_T$) distribution, and a
$\gamma\bar{\gamma}$ transverse-mass distribution peaked around $m_H$. 
This exotic signature has been recently analyzed for the first time 
in \cite{Gabrielli:2014oya}, in a model independent way.
In particular, a parton-level analysis at the 8-TeV LHC  has been performed 
for the main Higgs-boson production channel, namely  the gluon fusion process.
 Using the full 8-TeV LHC data set, a $5~\sigma$ sensitivity for a 
Higgs $H\!\to \!\gamma \, \bar{\gamma}$ branching ratio (BR) down to $0.5\%$
has been  obtained. These results have been worked out under  
assumptions that might underestimate one of the main reducible backgrounds, given by a photon plus jet ($j$), and did not include parton-shower effects.

The purpose of the present paper is twofold. On the one hand, we upgrade our previous 
8-TeV analysis 
of the $H \to \gamma \bar{\gamma}$ decay in the main Higgs production channel  
\cite{Gabrielli:2014oya} 
 by 
including parton-shower effects to the previous parton-level Montecarlo study of the signal and of  SM backgrounds. We also consider a more realistic background modeling, based on recent experimental studies of events with a photon plus missing energy 
at the LHC~\cite{Khachatryan:2015vta}. We then extend the analysis to the upgraded {\it nominal} LHC  energy 
of~14~TeV. 
On the other hand, we analyse for the first time an alternative signature coming from  
the $H \to \gamma \bar{\gamma}$ decay for a  Higgs boson produced via the Vector-Boson-Fusion (VBF) mechanism.
The gluon-fusion channel will turn out to be the most sensitive to  
BR($H \to \gamma \bar{\gamma}$) (BR$_{\gamma \bar{\gamma}}$). Nevertheless, we will see that the VBF process could significantly contribute to either a measurement or a determination of upper bounds of the decay 
rate of the exotic Higgs decay into a dark photon, possibly giving an independent confirmation of the signal in case of a positive observation in the gluon-fusion process.

The plan of the paper is the following.  In Sec.~II, we describe a theoretical
framework that might give rise to the $H\to\gamma \bar{\gamma}$ signature with observable rates. In Sec.~III ({A}), we study the potential of the gluon-fusion 
 Higgs production mechanism at the LHC 
for  constraining the $H \to \gamma \bar{\gamma}$ rate,
by a detailed analysis of both the signal and main backgrounds. The same is done for 
the VBF production mechanism  in Sec.~III ({B}).
In Sec.~IV, we summarize our results and conclude.

\section{Theoretical framework}
We now provide a model-independent parametrization of the amplitude
for the $H \to \gamma \bar{\gamma}$ channel, and then discuss the corresponding BR's range 
that can be expected in a class of NP models that might explain the origin and hierarchy of the Higgs Yukawa couplings.

The $H \to \gamma \bar{\gamma}$ amplitude  can be parametrized in a 
model-independent way by requiring gauge invariance, as follows
\bea
M_{\gamma\bar{\gamma}}&=& \frac{1}{\Lambda_{\gamma\bar{\gamma}} }\, T_{\mu\nu}(k_1,k_2) 
\varepsilon_1^{\mu}(k_1) \varepsilon_2^{\nu}(k_2),
\label{Mgg}
\eea
where $\Lambda_{\gamma\bar{\gamma}}$ is the effective scale associated
to the NP, 
$T^{\mu\nu}(k_1,k_2)\equiv g^{\mu\nu} k_1\cdot k_2 -k_2^{\mu}k_1^{\nu}\, $, 
and  $\varepsilon_{1}^{\mu}(k_{1})$  and $\varepsilon_{1}^{\mu}(k_{1})$ 
are the  photon and dark-photon polarization vectors, respectively.
The corresponding decay width is given by 
\bea
\Gamma(H\to \gamma \bar{\gamma})=m_H^3/(32\, \pi\, 
\Lambda_{\gamma\bar{\gamma}}^2).
\eea
A massless dark photon does not couple  to SM particles at tree level. One can 
 then assume that
 the effective amplitude in Eq.~(\ref{Mgg})
arises at one loop by the 
exchange  inside the loop of dark and messenger fields, the latter 
being charged under both SM and extra $U(1)_F$ gauge interactions.
By  naive dimensional analysis, one  expects  the   $\Lambda_{\gamma\bar{\gamma}}$ 
scale be proportional to the mass of the heaviest particle running in the loop, presumably related to the dark-sector.  If this were the case, the chances of observing this process at the LHC would be dramatically 
limited to a light dark-sector scenario, which is a quite strong requirement.
On the contrary, due to the non-decoupling properties  of the Higgs boson, this scale could be proportional to the Higgs vacuum expectation value (vev) 
(similarly to what happens for  the $H\to \gamma \gamma,\gamma Z,gg$ decay rates),
which would  allow for potentially large rates regardless of the characteristic mass scale of the dark sector. This crucial property turns out to hold 
in the framework of the model proposed in \cite{Gabrielli:2013jka}, as 
has been explicitly verified in \cite{Gabrielli:2014oya}. This framework can then  be used as a benchmark model for computing all the relevant quantities for predicting the Higgs decay rates into dark photons.

In \cite{Gabrielli:2013jka},
the Flavor and Chiral Symmetry breaking (ChSB) are assumed to 
be generated in a dark sector, and transferred to the Higgs Yukawa sector at one loop via Higgs-portal type scalar-messenger fields.
A new exact $U(1)_F$ gauge
symmetry in the dark sector produces  via a nonperturbative mechanism an 
exponential spread in the Yukawa couplings  $Y_i$ (with $i$ a flavor index), 
 providing 
a natural solution to the SM Flavor hierarchy problem.
Apart from the gauge boson of the unbroken $U(1)_F$ gauge group (the massless dark photon),
the dark sector consists of SM-singlet massive dark fermions, $Q_i$, a sort of
 rescaled replica of SM fermions. The requirement that the  gauge sector 
is unbroken allows dark fermions, which have $U(1)_F$ charges,  to be stable and thus potential dark-matter candidates. In addition to the dark sector, there are
scalar messenger fields (with the same quantum numbers as the
squarks and sleptons of supersymmetric models),  which communicate the ChSB and Flavor breaking from the dark sector to the Yukawa couplings. 

By restricting, for instance, only to the contribution of
colored messenger fields, 
the effective $\Lambda_{\gamma\bar{\gamma}}$ scale  can then be exactly derived 
 in the low energy limit  \cite{Gabrielli:2014oya}. 
In particular, for a universal average messenger mass $\bar{m}$, one obtains (up to corrections of order
${o}(m_H^2/\bar{m}^2)$)
\bea
\frac{1}{\Lambda_{\gamma\bar{\gamma}}} &=& 
\frac{R\sqrt{\alpha\bar{\alpha}}}{6\pi v}
\frac{\xi^2}{1-\xi^2}\, ,
\label{Lambda}
\eea
where $v$ is the  Higgs vev, \mbox{$R=N_c \sum_{i=1}^3\left(\eu\qui+\ed\qdi\right)$},  with $\qui, \qdi$ the $U(1)_F$ charges in the up and down sectors,
 $\eu=\frac{2}{3}$, $\ed=-\frac{1}{3}$ the corresponding e.m. charges,    
$\alpha$  the EM  fine structure constant,
and $N_c=3$ is the number of colors. Also, $\xi = \Delta/\bar{m}^2$,
with $\Delta=\mu_S v$ parametrising the left-right mixing of the messengers 
scalars, and $\mu_S$ is the vev of a singlet scalar field. The latter  spontaneously breaks  the  $H\to - H$ parity symmetry needed to 
forbid Higgs Yukawa interactions at tree level, since Yukawa couplings are 
generated radiatively  \cite{Gabrielli:2013jka}.

The non-decoupling
properties of the Higgs boson clearly show up in Eq.(\ref{Lambda}).
Indeed, the effective $\Lambda_{\gamma\bar{\gamma}}$ scale 
turns out to be proportional to the Higgs vev,  that is it tends
to a finite  value in the limit $\bar{m}\to \infty$ (for fixed  
mixing parameter $\xi <1$).  As stressed in 
\cite{Gabrielli:2014oya}, this is a general property of the Higgs boson, and does not depend on the peculiar structure of the model in \cite{Gabrielli:2013jka},
provided a messenger sector letting  the SM and the dark sector 
communicate exists.

The same off-shell fields contributing to the 
$H\to \gamma \bar{\gamma}$ decay amplitude at one loop can induce the $H\to \bar{\gamma} \bar{\gamma}$ 
transition to two dark photons (that increases the invisible Higgs decay width),
and also give  extra contributions to the $H\to \gamma \gamma,\gamma Z,gg\,$ 
SM decay rates. By parametrizing  these effects in a model independent
way,  BR$_{\gamma \bar{\gamma}}$ 
 values up to 5\% can be allowed, while
respecting all other LHC constraints~\cite{Gabrielli:2014oya}. 
Such large BR values for $H\to \gamma \bar{\gamma}$ are natural  in the framework of the model in  \cite{Gabrielli:2013jka} (see also~\cite{Biswas:2015sha} for further more model-dependent 
predictions). 

Such high decay rates strongly motivated the study of the Higgs production followed by 
the $H\to \gamma \bar{\gamma}$ decay at the  LHC Run-1 energy and integrated luminosity \cite{Gabrielli:2014oya}. The corresponding signature is indeed quite distinctive, 
with an almost monochromatic and massless  invisible (dark-photon) system and equally monochromatic photon, jointly resonating at the Higgs mass. 

In the present study, we will extend our previous analysis to the 14-TeV LHC setup,
upgrading different aspects of the study of the main gluon-fusion production channel, and including VBF Higgs production in order to improve the final  sensitivity to the 
$H\to \gamma \bar{\gamma}$ signature.

\section{Phenomenological analysis}

\subsection{Gluon-fusion channel}
We start by extending our previous LHC analysis  
of the gluon-fusion process at 8~TeV~\cite{Gabrielli:2014oya} 
to  14 TeV, improving the treatment of both the signal and  the main SM backgrounds. 
A crucial point in the refinement of  most important backgrounds 
will be the use of recently published experimental data by the CMS collaboration \cite{Khachatryan:2015vta}, where the relevant SM backgrounds are measured and reported. We will model our background accordingly. All this will result in a higher reliability of our signal and background estimates, that will anyhow substantially confirm our previous results on discovery potential based on a more {\it naive} analysis.

The process $pp\rightarrow H\rightarrow \gamma\bar{\gamma}$, where the Higgs is produced in the gluon-fusion channel, is characterized by a single photon recoiling against missing transverse momentum. In our previous analysis we outlined a search strategy for this process, based on the following  requirements (now slightly updated to take into account smearing effects discussed in the following):
\begin{itemize}
\item one isolated ($\Delta R > 0.4$) photon with $p_T^\gamma > 50\ {\rm GeV}$, 
and $|\eta^\gamma|<1.44$;
\item missing transverse momentum satisfying $\slashed{E}_T > 50$~GeV;
\item transverse mass in  the range $100\ {\rm GeV} < M^{T}_{\gamma\bar\gamma} < 130\ {\rm GeV}$;
\item no isolated leptons.
\end{itemize}
The transverse-mass variable  is defined as $M^{T}_{\gamma\bar\gamma}=\sqrt{2p_T^\gamma \slashed{E}_T(1-\cos\Delta\phi)}$, where   $\Delta\phi$ is the azimuthal distance between the photon transverse momentum $p_T^\gamma$, and the missing transverse momentum $\slashed{E}_T$. 

The main SM background for the above  selection criteria is $pp\rightarrow \!\gamma j$, where the missing transverse momentum can arise  from  a) neutrinos following heavy-flavor decays in the jet, b) mismeasurement of the jet energy, and c) very forward particles escaping the detector. To the latter channel contributes also  $p p \rightarrow j j$, whenever one of the jets is misidentified as a photon.
We assume the corresponding mis-tagging  probability  to be 0.1\%. Also, a photon identification efficiency of 90\% is adopted throughout this  analysis.
In our previous study~\cite{Gabrielli:2014oya}, the hadronic SM background was estimated at parton level  in a quite crude way, by treating any parton with $|\eta|>4.0$ as missing energy.

The CMS analysis of the data set at 8 TeV in \cite{Khachatryan:2015vta} assumes event selection criteria quite similar to the above,  in  order to search for an exotic  three-body decay of the Higgs boson into a photon and two invisible particles. Unfortunately, the CMS  analysis imposes an upper limit of 60 GeV on the photon transverse momentum, cutting away an important fraction of the signal region  for the two-body decay  of interest here
(for which $p_T^\gamma \lsim m_H/2$). However, due to the similarity of the residual event selection criteria in the two analysis, the continuous SM backgrounds are  expected to be comparable. As a consequence, we decided to model our  QCD background
according to the CMS measured distributions, benefitting from the highly optimized experimental procedure for the missing transverse-momentum determination.
This will  lead to a much improved reliability of our background estimate in the gluon-fusion channel.

We started by simulating the $\gamma j$ and dijet  backgrounds with 
MadGraph5\_aMC@NLO~(v2.2.2) \cite{Alwall:2014hca}, interfaced with PYTHIA~(v6.4.28) \cite{Sjostrand:2006za}, hence including  initial- and final-state radiation (ISR and FSR), hadronization and detector-resolution effects in the present updated analysis. We have generated event samples both at 8~TeV and 
14~TeV. We have then matched our  8-TeV samples to  the event yield corresponding to  the {\it 'SUSY benchmark'} event selection criteria reported in the CMS analysis~\cite{Khachatryan:2015vta}. 
This matching results in  $k$-factors connecting  our simulated samples to experimental data at 8 TeV.  We find $k=0.11$ for the $\gamma\, j$ background, and $k=0.058$ for the $j\!\rightarrow \!\gamma$ background. The order-of-magnitude reduction in the background estimate reported by  CMS as compared to our simulation is to be understood as a result of CMS advanced strategies for reducing event yields arising from mis-measured missing transverse momentum in hadronic events, as detailed in \cite{Khachatryan:2015vta}. It is beyond the scope of this work to attempt to exactly reproduce the CMS analysis. Instead, we assume that the CMS optimization  strategy works with comparable efficiency also in 14-TeV collisions, and that the corresponding reduction of the 14-TeV hadronic SM backgrounds is reliably captured by  rescaling our simulated samples with the same $k$ factors obtained from the  8-TeV matching.

We also upgraded the simulation of $H\rightarrow\gamma\bar{\gamma}$  signal events by including the ISR effects. Accordingly, 
we simulated Higgs production in association with either one or no jets with 
ALPGEN~(v2.14) \cite{Mangano:2002ea}, interfaced with PYTHIA  for jet-parton matching, hadronization and detector-resolution effects (see Sec. III (B) for the  jet definition and other simulation details). 

\begin{figure}
\centering
\includegraphics[width=0.47\textwidth]{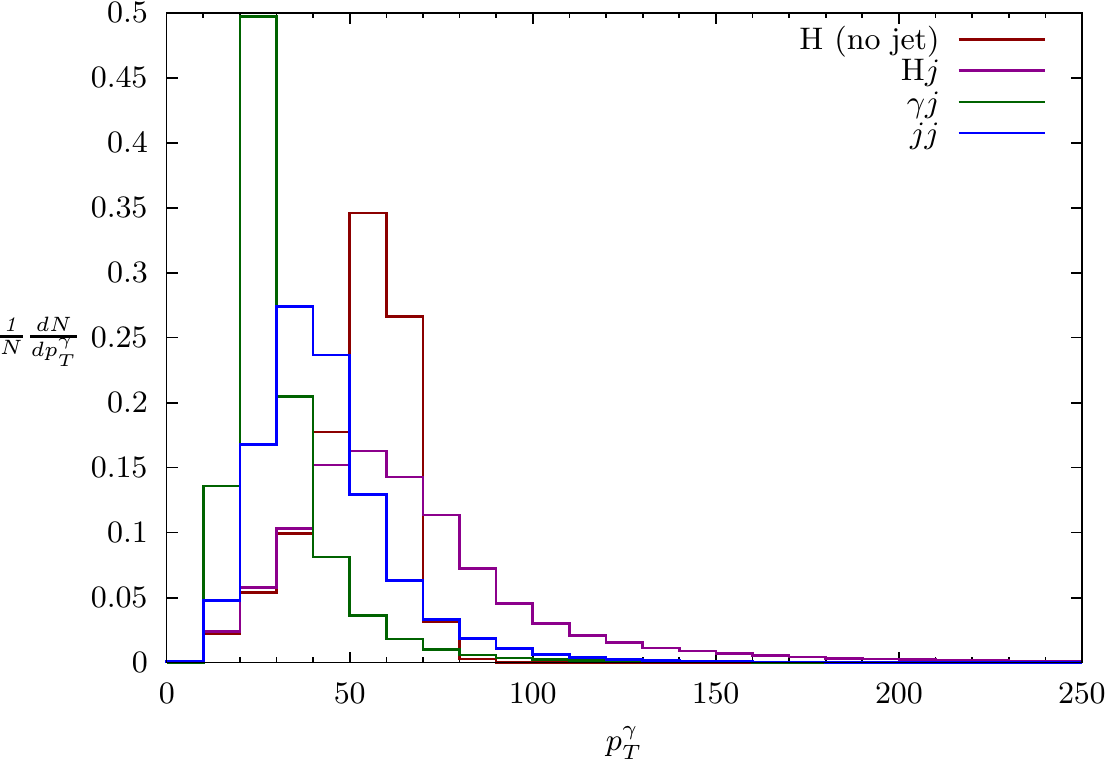}
\includegraphics[width=0.47\textwidth]{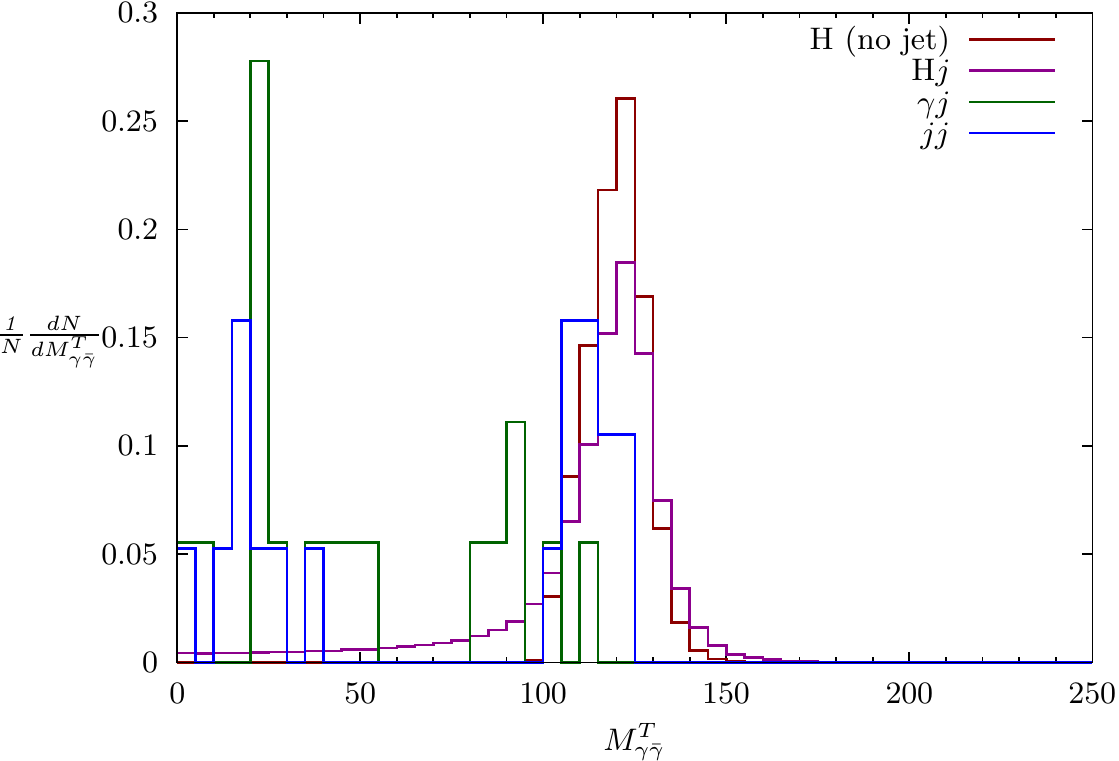}
\caption{Photon $p_T$ (upper plot) and  transverse-mass (lower plot) distributions for the $H\!\to \!\gamma\bar{\gamma}\,$ signal in the gluon-fusion process, and for SM backgrounds, for  
 inclusive $\gamma+\slashed{E}_T$ final states with no isolated leptons.  The effect 
of extra radiation on the signal events is also depicted. All distributions are normalized to unity.}
\label{fig:ggf_pt}
\end{figure}

The corresponding smearing in the  $p^\gamma_T$ and  $M^{T}_{\gamma\bar\gamma}$  spectra   for the $H\rightarrow\gamma\bar{\gamma}$ signal  is shown in Figure~\ref{fig:ggf_pt}. There, the two categories corresponding to no extra jets and one extra jet accompanying the Higgs signal are shown separatly, along with the distributions for the hadronic backgrounds coming from 
$\gamma\, j$ production, and dijet production followed by   $j\!\rightarrow \!\gamma$ mistagging. The latter distributions are obtained with a nominal cut 
on the photon transverse momentum, $p_T^\gamma > 10\ {\rm GeV}$, and $p_T^j > 10\ {\rm GeV}$ on fake jet in the dijet analysis.

Besause of  initial-state-radiation and detector-resolution effects, a better sensitivity for the signal is obtained by relaxing the maximum value of the photon
transverse-momentum cut, and increasing the transverse mass window from $100\ {\rm GeV} < M^{T}_{\gamma\bar\gamma} < 126\ {\rm GeV}$ to $100\ {\rm GeV} < M^{T}_{\gamma\bar\gamma} < 130\ {\rm GeV}$ with respect to~\cite{Gabrielli:2014oya}.

The main electroweak background consists of the channels $pp \rightarrow W \rightarrow e\nu$, where the electron is misidentified as a photon, $pp\rightarrow W(\rightarrow \ell\nu) \gamma$, for $\ell$ outside charged-lepton acceptance, and $pp\rightarrow Z(\rightarrow \nu\nu) \gamma$. We have simulated these processes at parton level according to the analysis in \cite{Gabrielli:2014oya} , using a $e\!\to\!\gamma$ conversion probability of 0.005 for the first process. 

In Table \ref{ggf}, one can find a summary of 
the cross sections times acceptance (in fb) for the signal and backgrounds at 8 TeV and 14 TeV for the gluon-fusion process, assuming BR$_{\gamma \bar{\gamma}}$=1\%, and obtained as discussed above.
\begin{table}
\begin{center}
\begin{tabular}{c|c|c}
 & $\sigma\times A$ \small{ [8 \!TeV]} & $\sigma\times A$\small{ [14 \!TeV]} \\ \hline
$H\!\to\! \gamma \bar{\gamma}\;\;$\small{ (BR$_{\gamma \bar{\gamma}}=1\%)$} & 44 & 101 \\ \hline
 $\gamma j$ & 63 & 202 \\
 $jj\rightarrow\gamma j$ & 59 & 432 \\
 $e\rightarrow \gamma$ & 55 & 93 \\
 $W(\rightarrow \!\ell\nu) \gamma$ & 58 & 123 \\
 $Z(\rightarrow \!\nu\nu) \gamma$ & 102 & 174 \\ \hline
\small{total background} & 337 & 1024 \\ \hline
\end{tabular}
\caption{Cross section times acceptance $A$ (in fb) for the gluon-fusion signal and backgrounds at 8 and 14 TeV, assuming BR$_{\gamma \bar{\gamma}}\!\!\!=\,$1\%, with the selection \mbox{$p_T^\gamma > 50\ {\rm GeV}$, $|\eta^\gamma|<1.44$}, \mbox{$\slashed{E}_T > 50$ GeV}, and \mbox{$100\ {\rm GeV} < M^{T}_{\gamma\bar\gamma} < 130\ {\rm GeV}$}.}
\label{ggf}
\end{center}
\end{table}

With the 20 fb$^{-1}$ data set at 8~TeV, our improved analysis gives  a $5\sigma$ discovery reach at BR$_{\gamma \bar{\gamma}}\simeq 4.8\times 10^{-3}$, compatible with our previous estimate~\cite{Gabrielli:2014oya}. The present more-realistic event simulation  was expected 
to deteriorate the capability of separating signal from background. This effect has been actually mostly compensated by the advanced optimization experimental strategies recently applied to  the missing transverse-momentum data, on which we have now modeled our background simulation.

Assuming an integrated luminosity of 100 (300) fb$^{-1}$ at 14 TeV, and extrapolating the effect of these optimization technique to higher energies, we find a $5\sigma$ discovery potential for  BR$_{\gamma \bar{\gamma}}$ down to $1.6\times 10^{-3}$($9.2\times10^{-4}$).  At the  High-Luminosity LHC (HL-LHC), with an integrated luminosity of 3 ab$^{-1}$,  the $5\sigma$ reach is extended down to $2.9\times 10^{-4}$.

\subsection{VBF channel}
We now turn our focus on the Higgs production in the VBF channel. This presents a lower production rate with respect to the gluon-fusion channel. On the other hand, it is in principle more controllable due to its strong kinematical characterization. In particular, the process  $pp \rightarrow Hjj\rightarrow\gamma\bar{\gamma}jj$, where the Higgs boson arises from
a $W(Z)$-pair fusion, results mostly in two forward jets with opposite rapidity, one photon and missing transverse momentum. 

We started by simulating the signal by PYTHIA, by including both the Higgs VBF production and its subsequent decay into a $\gamma\bar{\gamma}$ final state. 
 The main SM backgrounds are given by the production of QCD multi-jets, $\gamma+$jets, and 
 $\gamma+Z(\to\!\bar\nu \nu)+$jets. The $\gamma+$jets background 
has been simulated using  ALPGEN. We have generated $\gamma j$, $\gamma j j$, and $\gamma j j j$ samples  with $p_T^\gamma > 10$ GeV 
and $|\eta^\gamma|<2.5$ for photons, and $p_T^j>20$ GeV and $|\eta^j|<5$ for jets. An isolation of $\Delta R > 0.4$ between all pairs of objects is required.    
We have then interfaced ALPGEN and PYTHIA, and  incorporated the jet-parton matching, 
according to the MLM prescription \cite{Hoche:2006ph}. Events containing hard partons are generated  in  
ALPGEN  with a  cut on the transverse momentum ($p_T>20$ GeV), and 
on the rapidity ($|\eta|<5.0$) of each parton, along with a minimum separation ($\Delta R>0.4$) between 
them. These events are then interfaced with PYTHIA for showering, to take into account soft and 
collinear emission of partons. All partons are then clustered using a cone jet algorithm with   
$p_T>20$ GeV, and a cone size of $\Delta R=0.6$ (the latter used only for matching purposes, not 
for the jet definition in the event selection). An event is said to be matched if there is a one-to-one 
correspondence between jets and initial hard partons. An event with an extra jet which is not matched to a parton 
is rejected in case of exclusive matching, while is kept in case of inclusive matching for 
the highest jet-multiplicity samples.

For the QCD multi-jet process and 
the $\gamma+Z+$ jets process we have used MadGraph 5 interfaced with PYTHIA. In case of the QCD multi-jet process, the most central jet is assumed to be mistagged 
 as a photon with a corresponding faking probability of 0.1\%. The ISR and FSR effects, parton shower, hadronisation 
and finite detector resolution effects have also been implemented for  the signal and all backgrounds. We have then assumed a photon identification efficiency 
of 90\%. The distributions are obtained with a nominal cut on the photon 
transverse momentum, $p_T^\gamma > 10\ {\rm GeV}$, and $p_T^j > 10\ {\rm GeV}$ on fake jet in the QCD multijets analysis.

\begin{figure}
\centering
\includegraphics[width=0.47\textwidth]{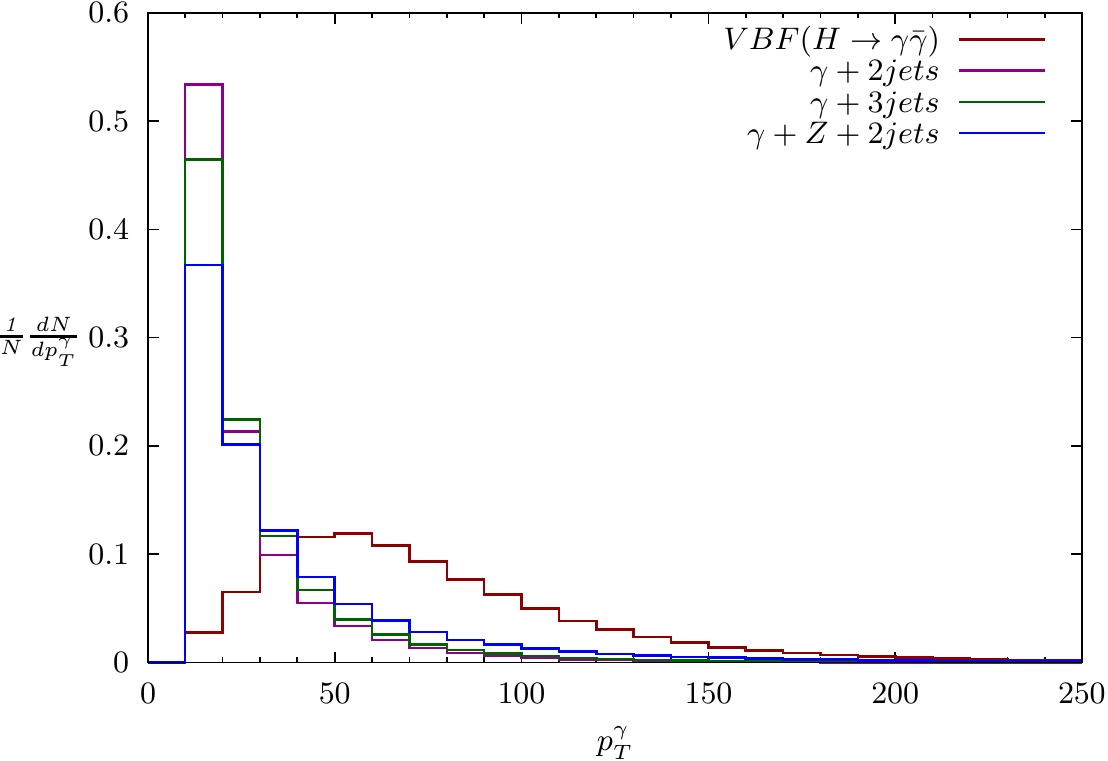}
\includegraphics[width=0.47\textwidth]{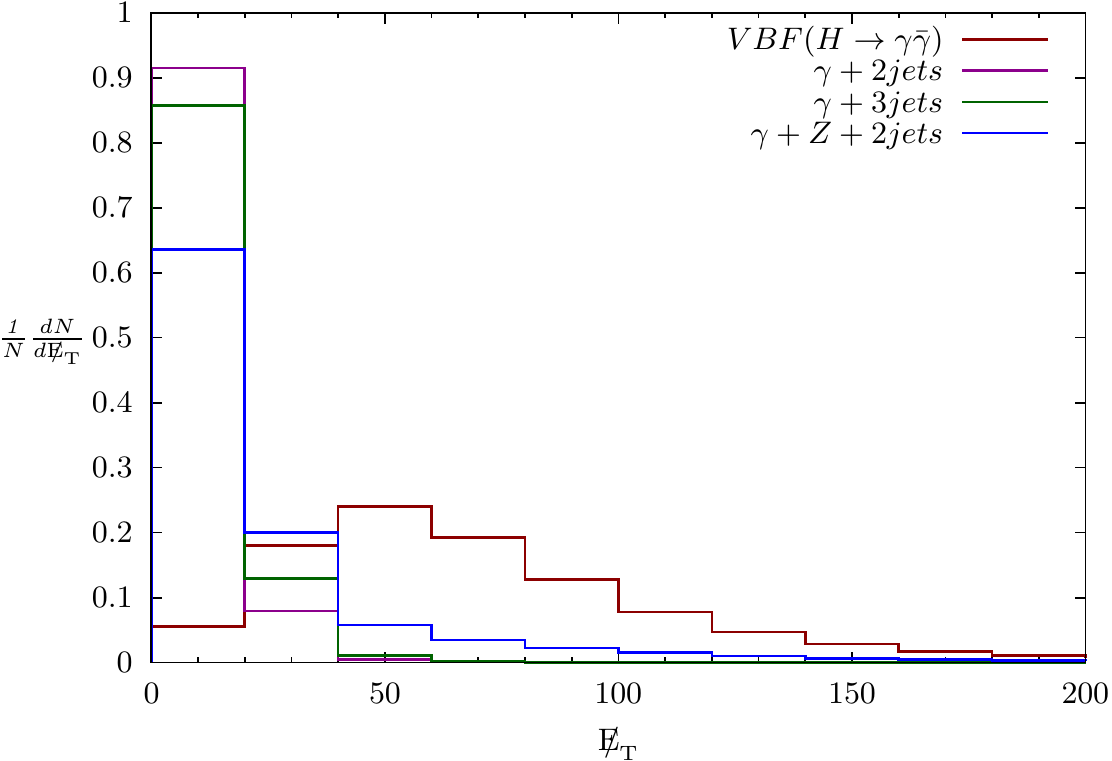}
\caption{ Photon $p_T$ (upper plot), and missing transverse-energy (lower plot) distributions for the signal and SM backgrounds in the VBF process. The 
final state in this case is $\gamma+\slashed{E}_T+(\geq \!2)$jets with no isolated leptons.  All distributions are 
normalized to unity.}
\label{fig:vbf1}
\end{figure}

\begin{figure}
\centering
\includegraphics[width=0.47\textwidth]{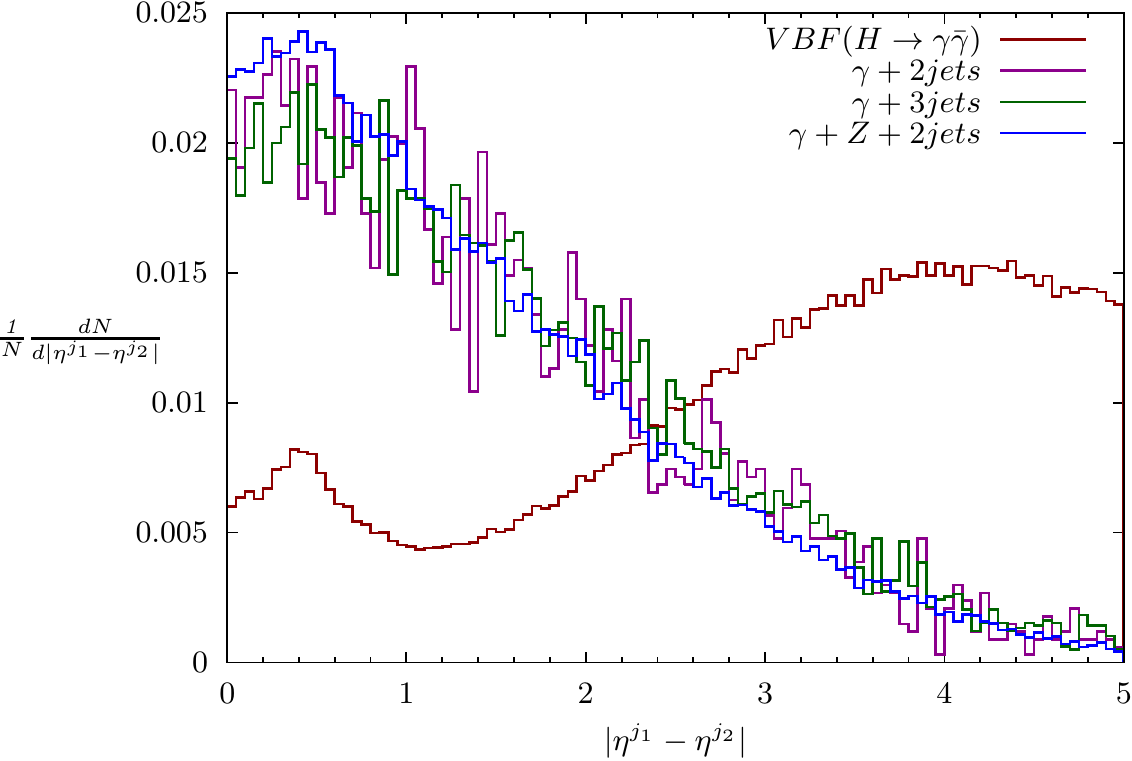}
\includegraphics[width=0.47\textwidth]{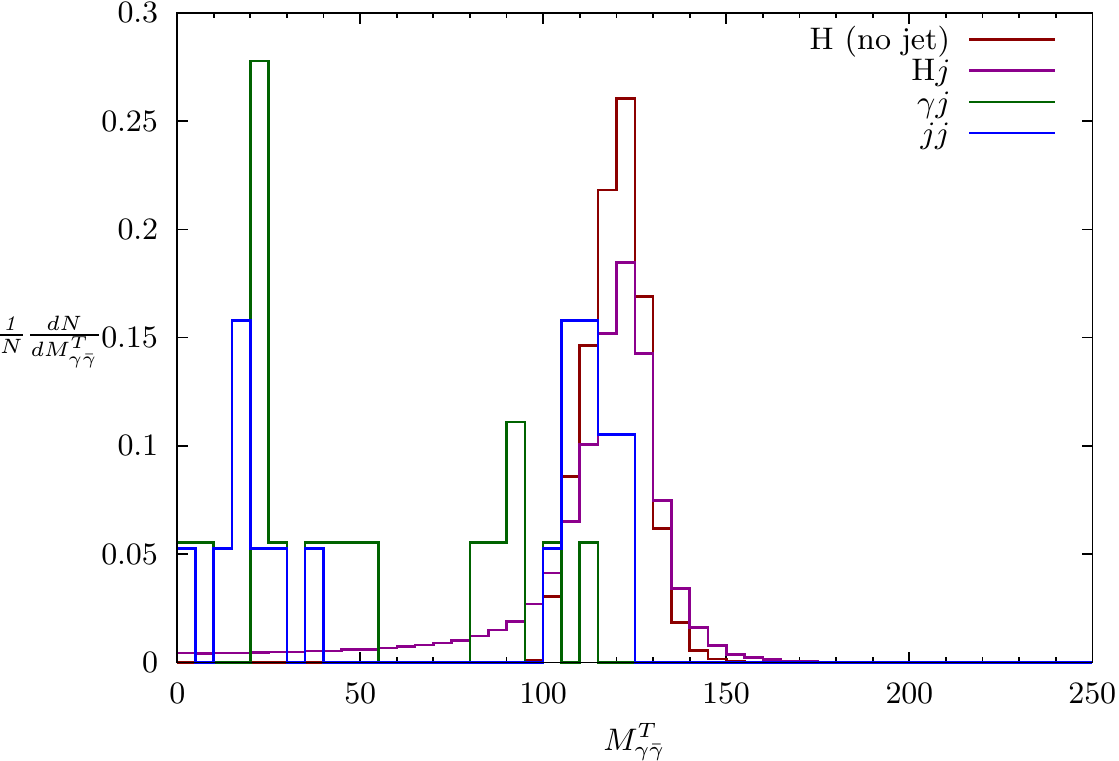}
\caption{Rapidity gap between the two forward jets (upper plot),  and  transverse-mass (lower plot)  distributions for the signal and 
 SM backgrounds in the $\gamma+\slashed{E}_T+\!(\geq 2)$jets final state with no isolated leptons. The $\Delta\eta = |\eta^{j_1}-\eta^{j_2}|$ distribution is 
obtained with a cut  $p_T^\gamma > 30\ {\rm GeV}$,  for $p_T^j > 30\ {\rm GeV}$ on the fake jet in the QCD multijets 
analysis, and $\slashed{E}_T > 30$ GeV. The transverse mass distribution is obtained with the additional cuts \mbox{$\eta^{j_1}\times\eta^{j_2}<0$} and 
$|\eta^{j_1}-\eta^{j_2}|>4.0$. All distributions are normalized to unity.}
\label{fig:vbf2}
\end{figure}

In Figures \ref{fig:vbf1} and \ref{fig:vbf2}, we plot a few kinematic distributions which are  useful to separate the signal from the backgrounds.


On this basis, we propose to select the events according to the following criteria:

\begin{itemize}
\item ({\it basic cuts}) one isolated photon with $p_T^\gamma > 30$ GeV and $|\eta^\gamma|<2.5$, and two or more jets with $p_T^j>20$ GeV and $|\eta^j|<5.0$, and 
angular separation $\Delta R > 0.4$ between all objects;

\item ({\it basic cut}) missing transverse energy $\slashed{E}_T > 30$ GeV;

\item ({\it basic cut}) no isolated leptons;

\item ({\it rapidity cuts}) rapidities of the two highest $p_T$ jets  obey \mbox{$\eta^{j_1}\times\eta^{j_2}<0$} and \mbox{$|\eta^{j_1}-\eta^{j_2}|>4.0$};

\item ($M^{T}_{\gamma\bar\gamma} $ {\it cuts}) transverse mass of the photon and invisible system satisfying $100 {\rm ~GeV}<M^{T}_{\gamma\bar\gamma} < 130$ GeV (as above, the upper bound has been extended with respect to $m_H$ to take into account the smearing of the $M^{T}_{\gamma\bar\gamma}$ distribution, {\it cf.} Figure~\ref{fig:vbf2}).

\end{itemize}

In Table~\ref{vbf1}, we present the cross sections for the signal and dominant SM backgrounds after the sequential application of  basic cuts,
rapidity cuts on the two forward jets, and  transverse-mass cut on the photon plus missing transverse-energy system.

In order to better control the missing transverse energy arising from jet energy mis-measurements, we have also imposed an azimuthal  isolation cut
$\Delta\phi(j_i,\slashed{E}_T) >$ 1.5 (with $i=1,2$) on the angles between the ${\slashed{E}_T}$
direction  and the transverse momenta of the two 
highest-$p_T$ jets. 

Furthermore, we studied the effect of a selection cut occasionally applied for searches in the VBF channel (see, \eg, the $W\to \ell\nu$ analysis in VBF in \cite{CMS:2015uta}).
This is  the $y^*<1.0$ cut  on the Zeppenfeld 
variable  defined as $y^* = |y^H -\frac12(\eta^{j_1}-\eta^{j_2})|$, where the Higgs rapidity  $y^H$ is reconstructed from the photon momentum and the 
missing transverse energy as described in~\cite{Rainwater:1996ud}. $X$ systems produced 
via  
VBF  are in fact characterized by a smaller $y^*$ value, with respect to
other $X$+2-jet backgrounds.
The values of the $\Delta\phi(j_i,\slashed{E}_T)$ and  $y^*$  cuts have been  separately optimized in order to  increase the signal significance.

Table~\ref{vbf2} presents the independent effect of the  $y^*$ and $\Delta\phi(j_i,\slashed{E}_T)$  cuts, applied after the set of 
cuts listed in Table~\ref{vbf1}.  The combined effect of these two cuts is also shown in the last row of 
Table~\ref{vbf2}. 
The $\Delta\phi(j_i,\slashed{E}_T)$ cut turns out to be much more effective in separating the signal from background. We then dropped the $y^*$ cut in our final selection.

Since the $\Delta\phi(j_i,\slashed{E}_T)$ distribution is asymmetric in the exchange of the first and second highest-$p_T$ jets, we have also 
 tried to optimize the signal significance by assuming an asymmetric cut on $\Delta\phi(j_i,\slashed{E}_T)$, that is by applying different cuts on the first and second highest-$p_T$ jets. We anyway found that the best signal to background ratio is obtained with the symmetric cut  $\Delta\phi(j_i,\slashed{E}_T) >$ 1.5 on both jets.

Finally, assuming an integrated luminosity of 300 fb$^{-1}$, in the last column of Table~\ref{vbf2} we present the estimated VBF signal significances for BR$_{\gamma \bar{\gamma}}$=1\%.    For this setup, 
the signal significance $S/\sqrt{S+B}$ approaches the 5$\sigma$ level. 
For  100 fb$^{-1}$, the 
5$\sigma$ reach  in branching ratio is  about BR$_{\gamma \bar{\gamma}}\!\!\simeq$ 2\%. 
With the HL-LHC integrated luminosity of 3 ab$^{-1}$, the 5$\sigma$ reach can be extended down to BR$_{\gamma \bar{\gamma}}\!=3.4\times 10^{-3}$.


\begin{table}[t]\scriptsize
\begin{center}
\begin{tabular}{l|c|c|c|c}
Cuts                           & Signal   &  $\gamma+$jets   &  $\gamma+Z+$jets  &  QCD multiijet  \\ \hline
Basic cuts                    &  17.7    &  266636               &              1211            &   72219            \\
Rapidity cuts                &    8.8     &   8130                  &               38.1            &   33022             \\
$M^{T}_{\gamma\bar\gamma}$ cuts         &    5.0     &     574                  &                6.5             &     3236             \\
\hline
\end{tabular}
\end{center}
\caption{Cross sections  times acceptance $\sigma\times A$ (in fb) for  the VBF signal and backgrounds at 
14~TeV, after sequential application of cuts defined in the text, assuming BR$_{\gamma \bar{\gamma}}$=1\%.}
\label{vbf1}
\end{table}

\begin{table}[t]\scriptsize
\begin{center}
\begin{tabular}{l|c|c|c|c|c}
Cuts                                                                      & Signal    &  $\gamma+$jets   &  $\gamma+Z+$jets   & multijet & L=300 fb$^{-1}$  \\ 
\hline
$y^*<1.0$                                                        &  2.67       &       84.2           &    1.84       &      758   &  1.6$\,\sigma$     \\
\hline
$\Delta\phi(j_i,\slashed{E}_T) >$1.5    &     1.82           &       6.9         &    2.16     &      37    & 4.6$\,\sigma$     \\
\hline \hline
both cuts     &     1.21     &       1.2     &   0.67    &     19  &  4.5$\,\sigma$    \\
\hline
\end{tabular}
\end{center}
\caption{Cross sections  times acceptance $\sigma\times A$ (in fb) for  the VBF signal and backgrounds at 
14~TeV,  assuming BR$_{\gamma \bar{\gamma}}$=1\%. The first and second row corresponds to the separate effect of the 
$y^*$ and $\Delta\phi(j_i,\slashed{E}_T)$ cuts, respectively, after applying all the cut sequence 
in Table~\ref{vbf1}. The last row represents the combined effects of the two cuts. The last column shows the signal significance  for an integrated luminosity of \mbox{L=300 fb$^{-1}$}. }
\label{vbf2}
\end{table}

\section{Summary and Conclusions}
We have studied the prospects for discovering an exotic Higgs-boson decay  into a SM photon and a new neutral massless vector boson, a dark photon, at the LHC with $\sqrt S=14$ TeV. We have updated our previous analysis of  the gluon-fusion channel at 8 TeV by a more reliable treatment of both the signal and hadronic SM backgrounds, and extended this approach to 14-TeV collisions. We also explored for the first time the possibility of detecting the exotic $H\!\to \!\gamma \, \bar{\gamma}$ channel in the VBF Higgs production. 

\begin{table}
\begin{tabular}{c|c c|c c|c c}
 BR$_{\gamma \bar{\gamma}}$ (\%)  &\multicolumn{2}{|c|}{\;L=100\,fb$^{-1}$} & \multicolumn{2}{|c|}{\;L=300\,fb$^{-1}$} &\multicolumn{2}{|c}{\;L=3\,ab$^{-1}$} \\ \hline
{\small Significance} & 3$\sigma$ & 5$\sigma$ & 3$\sigma$ & 5$\sigma$ & 3$\sigma$ & 5$\sigma$ \\ \hline
\;BR$_{\gamma \bar{\gamma}}$(VBF) &  1.1 & 1.9 &  0.65 & 1.1  &  0.21 & 0.34 \\ \hline
BR$_{\gamma \bar{\gamma}}\,$($ggF$) & \, 0.096 &  0.16   & \, 0.055  &  0.092  &\,  0.017  &  0.029  \\ \hline
\end{tabular}
\caption{Reach in BR$_{\gamma \bar{\gamma}}$ (in percentage) for a 3$\sigma$ exclusion or a 5$\sigma$ discovery at the 14 TeV LHC, in the VBF and 
gluon-fusion channels, for different integrated luminosities L.}
\label{summarytable}
\end{table}

A summary of our findings is presented in Table~\ref{summarytable}, where we show the predicted reach in detectable BR$_{\gamma \bar{\gamma}}$  for both  exclusion (at a 3$\sigma$ level) and discovery (at a 5$\sigma$ level), assuming  100, 300 and 3000 fb$^{-1}$ of  data at 14~TeV. The gluon-fusion potential turns out to be definitely higher, extending the  BR$_{\gamma \bar{\gamma}}$ reach with respect to the VBF channel by more than one order of magnitude. In particular, according to the present analysis, the full LHC program will allow to discover (exclude) a 
BR$_{\gamma \bar{\gamma}}$ value down to less than $1\times10^{-3} \,(6\times 10^{-4})$, while the HL-LHC phase  will be sensitive to  BR$_{\gamma \bar{\gamma}}$ as small as  $3\times 10^{-4}\,(2\times 10^{-4})$.
We recall that BR$_{\gamma \bar{\gamma}}$ 
 values up to 5\% are allowed in realistic BSM frameworks~\cite{Gabrielli:2014oya}.

In light of the projected discovery reach and of the theoretical interest in dark-photon models, we urge the ATLAS and CMS  experiments to perform a dedicated analysis of the $H\!\rightarrow\! \gamma+\slashed{E}_T$ signature in two-body final states. The event selection criteria used in the CMS analysis \cite{Khachatryan:2015vta}, by imposing an upper limit of 60 GeV on $p_T^\gamma$,  considerably restrict the signal phase space for the two-body decay mode. Nevertheless, the methods used by CMS for the suppression of the SM hadronic backgrounds to the $\slashed{E}_T$ signature can be very effective even for  relatively low transverse-momentum final states, possibly resulting in experimental sensitivities for branching ratios well below the permil
 level. Similar methods could actually be applied (once the corresponding experimental analyzes will be available) for suppressing the SM multi-jet background to the VBF channel, possibly increasing the relative weight of the VBF analysis in the search for a $H\to\gamma \bar{\gamma}$ signature, hence expanding the LHC potential.

After the recent observation at the LHC of an excess in the di-photon spectrum around an invariant mass of about 750 GeV~\cite{ATLAS750,CMS:2015dxe}, it would be also advisable to  extend the search for $\gamma+\slashed{E}_T$ final states to higher invariant masses of the $\gamma \bar{\gamma}$ pair. Indeed, 
the observed features of the would-be 750-GeV $\gamma{\gamma}$ resonance might require new degrees of freedom in a hidden sector in order to give rise to effective couplings to photons (and gluons) (see,\eg, \cite{Franceschini:2015kwy}). The latter  degrees of freedom could well be portals to a massless dark photon, in case they are also charged under an extra unbroken $U(1)_F$. Since a large $U(1)_F$ coupling might be naturally allowed~\cite{Biswas:2015sha}, the corresponding rate for a $\gamma \bar{\gamma}$ resonance at 750 GeV could already be sizable with the present data set. This possibility has also  been envisaged in~\cite{Tsai:2016lfg}.

In case the di-photon signature will be confirmed at the LHC, the search for  new structures in the $\gamma+\slashed{E}_T$ transverse-mass distributions at 750 GeV would provide extra invaluable insight about the nature of the NP behind it.

\mysection{Acknowledgments}
We thank Daniel Fournier,  Jean-Baptiste de Vivie de R\'egie, and Rachid Mazini for useful discussions.
E.G. would like to thank the TH division of CERN for its kind hospitality during the preparation of this work.
The work of M.H. has been supported by the Academy of Finland project number 267842.

\end{document}